\pdfoutput=1
\documentclass[11pt]{article}
\usepackage[ruled,vlined]{algorithm2e}
\usepackage{acl}
\usepackage{times}
\usepackage{latexsym}

\usepackage[T1]{fontenc}
\usepackage[utf8]{inputenc}
\usepackage{amssymb}

\usepackage{microtype}
\usepackage{pifont}

\usepackage{inconsolata}

\usepackage{graphicx}

\usepackage{amsmath}
\usepackage{amssymb}
\usepackage{bm}

\usepackage{amsfonts}
\usepackage{amssymb}

\usepackage{algpseudocode}
\usepackage{times}
\usepackage{latexsym}
\usepackage{soul}
\usepackage[T1]{fontenc}
\usepackage{verbatim}
\usepackage{booktabs}
\usepackage{hyperref}
\usepackage{multirow}
\usepackage[table]{colortbl}
\usepackage{graphicx}
\usepackage{amsmath}
\usepackage{caption}
\usepackage{subcaption}
\usepackage{tabularx}
\usepackage{url}
\usepackage{xurl}
\usepackage[most]{tcolorbox}
\usepackage{amsthm}

\theoremstyle{plain}
\newtheorem{theorem}{Theorem}[section]
\newtheorem{lemma}[theorem]{Lemma}

\theoremstyle{definition}

\theoremstyle{remark}

\definecolor{LavenderBlush}{rgb}{1.0, 0.94, 0.96}
\definecolor{Lavender}{rgb}{0.9, 0.9, 0.98}
\definecolor{MistyRose}{rgb}{1.0, 0.89, 0.88}
\definecolor{MintCream}{rgb}{0.96, 1.0, 0.98}
\definecolor{AliceBlue}{rgb}{0.94, 0.97, 1.0}
\definecolor{Seashell}{rgb}{1.0, 0.96, 0.93}
\definecolor{LightYellow}{rgb}{1.0, 1.0, 0.88}
\definecolor{Peach}{rgb}{1.0, 0.9, 0.71}
\definecolor{Apricot}{rgb}{0.98, 0.81, 0.69}
\definecolor{LightCyan}{rgb}{0.88, 1.0, 1.0}
\definecolor{rose}{rgb}{1.0, 0.0, 0.5}
\definecolor{mint}{rgb}{0.74, 0.99, 0.79}
\definecolor{coral}{rgb}{1.0, 0.5, 0.31}

\definecolor{PeachPuff}{rgb}{1.0, 0.85, 0.73}
\definecolor{Beige}{rgb}{0.96, 0.96, 0.86}
\definecolor{LightSalmon}{rgb}{1.0, 0.63, 0.48}
\definecolor{Ivory}{rgb}{1.0, 1.0, 0.94}
\definecolor{MintCream}{rgb}{0.96, 1.0, 0.98}

\definecolor{Red}{rgb}{1.0, 0.0, 0.0}
\definecolor{Green}{rgb}{0.0, 0.5, 0.0}
\definecolor{Blue}{rgb}{0.0, 0.0, 1.0}
\definecolor{Orange}{rgb}{1.0, 0.55, 0.0}
\definecolor{Purple}{rgb}{0.5, 0.0, 0.5}
\definecolor{Goldenrod}{rgb}{0.85, 0.65, 0.13}
\definecolor{BurntOrange}{HTML}{CC5500}
\definecolor{Cyan}{rgb}{0.0, 1.0, 1.0}
\definecolor{Maroon}{rgb}{0.5, 0.0, 0.0}  % Standard maroon RGB

\definecolor{textblue}{RGB}{25,25,112}      % MidnightBlue variant
\definecolor{textred}{RGB}{139,0,0}         % DarkRed variant
\definecolor{Pink}{RGB}{255,192,203}  % defines 'Pink' as RGB(255,192,203)
\definecolor{softgray}{RGB}{220,220,220}
\definecolor{softblue}{RGB}{230,245,255}
\definecolor{softred}{RGB}{255,235,238}
\definecolor{framegray}{RGB}{100,100,100}
\definecolor{frameblue}{RGB}{100,149,237}   % CornflowerBlue
\definecolor{framered}{RGB}{220,20,60}      % Crimson

\definecolor{LightGreen}{rgb}{0.88, 1.0, 0.88}
\definecolor{LightPink}{rgb}{1.0, 0.9, 0.9}
\definecolor{LemonChiffon}{rgb}{1.0, 0.98, 0.8}

\definecolor{Brown}{rgb}{0.65, 0.16, 0.16}
\definecolor{LightSalmon}{rgb}{1.0, 0.63, 0.48}  % If not defined earlier
\definecolor{Gray}{rgb}{0.5, 0.5, 0.5}           % 60% gray approximation
\definecolor{DarkGreen}{rgb}{0.0, 0.39, 0.0}

\newcommand{\cross}{\ding{55}} % Cross mark

\title{Can Argus Judge Them All? Comparing VLMs Across Domains}

\author{
\textbf{Harsh Joshi}\textsuperscript{1},
\textbf{Gautam Siddharth Kashyap}\textsuperscript{2},
\textbf{Rafiq Ali}\textsuperscript{3},
\textbf{Ebad Shabbir}\textsuperscript{3},
\textbf{Niharika Jain}\textsuperscript{4},\\
\textbf{Sarthak Jain}\textsuperscript{5},
\textbf{Jiechao Gao}\textsuperscript{6}\thanks{Corresponding Author: jiechao@stanford.edu},
\textbf{Usman Naseem}\textsuperscript{2} \\
\textsuperscript{1}Bharati Vidyapeeth, New Delhi, India \\
\textsuperscript{2}Macquarie University, Sydney, Australia \\
\textsuperscript{3}DSEU-Okhla, New Delhi, India \\
\textsuperscript{4}Vivekananda Institute of Professional Studies, New Delhi, India \\
\textsuperscript{5}IIIT-Delhi, India \\
\textsuperscript{6}Center for SDGC, Stanford University, California, USA \\
}

\begin{document}
\maketitle
\begin{abstract}
Vision-Language Models (VLMs) are increasingly used in industry VLM applications such as retrieval systems, content generation platforms, and decision-support workflows, where model selection is commonly guided by benchmark rankings. These rankings are largely determined by retrieval, captioning, and reasoning downstream tasks; however, models with similar task performance often show substantially different behavior across datasets. This creates a Capability-Reliability Gap between benchmark performance and observed model stability. We present ARGUS-EVAL, a capability-reliability-oriented evaluation framework for VLMs that characterizes model behavior through Benchmark Capability $P(M)$, Cross-Dataset Consistency $CDC(M)$, Robustness Retention $RR(M)$, and Efficiency $E(M)$. We evaluate CLIP, BLIP, LXMERT, Gemma-3-4B, and Qwen-2.5VL-3B-Instruct across retrieval, captioning, and reasoning downstream tasks. The results reveal notable differences between capability-oriented and reliability-oriented rankings. Qwen-2.5VL-3B-Instruct achieves the strongest overall capability ($R@1=82.7\%$, $BLEU\text{-}4=47.2\%$, $CIDEr=141.6$, $CDC=0.91$), whereas CLIP records the lowest latency ($31\,ms$) and memory footprint ($0.9\,GB$).
\end{abstract}

\begin{comment}
\begin{table}[t!]
\centering
\tiny
\begin{tabular}{l|c}
\toprule
\textbf{Model} & \textbf{CDC Score ($\uparrow$)} \\
\midrule
LXMERT & 0.64 \\
BLIP   & 0.76 \\
Gemma-3-4B & 0.85 \\
Qwen-2.5VL-3B-Instruct & \textbf{0.91} \\
CLIP   & 0.88 \\
\bottomrule
\end{tabular}
\caption{Cross-Dataset Consistency (CDC) scores for five VLMs. Higher CDC indicates greater robustness and performance stability across heterogeneous datasets. Instruction-tuned models, particularly Qwen-2.5VL-3B-Instruct, exhibit the highest CDC values, reflecting superior generalization and reduced domain variance. Bold values denote the best results in each column.
}
\label{tab:cdc-results}
\end{table}
\end{comment}

\begin{figure}[t]
\vspace{-0.3cm}
    \centering
    \includegraphics[width=0.85\linewidth]{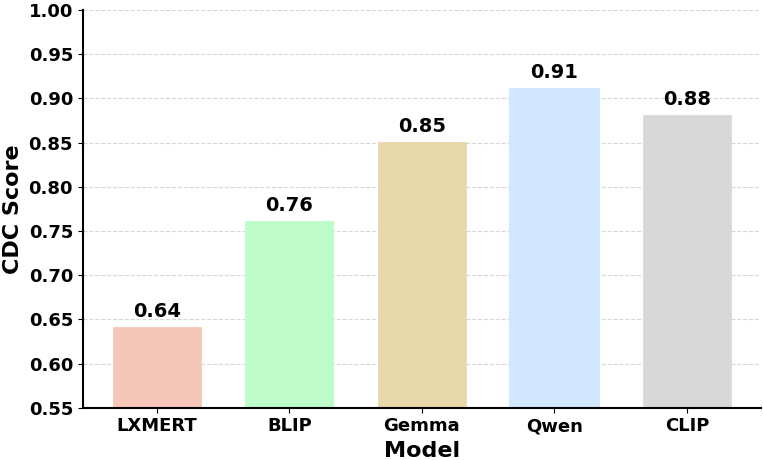}
    \vspace{-0.2cm}
    \caption{
Illustration of the Capability-Reliability Gap in industry VLM applications. Despite achieving benchmark capability above $80\%$, representative VLMs show up to $23.4\%$ variation in cross-dataset performance stability, resulting in substantially different $CDC(M)$ scores (see \S\ref{Downstream Analysis}).
}
    \label{fig:capability_reliability_gap}
    \vspace{-0.4cm}
\end{figure}

\section{Introduction}
\label{Introduction}

Vision-Language Models (VLMs) combine visual perception and language understanding within a unified architecture for machines to retrieve \cite{singh2026vision}, caption \cite{liang2026comprehensive}, and reason \cite{nguyen2026vlm} about multimodal content \cite{ma2026survey}. As VLMs become increasingly integrated into retrieval systems, content generation platforms, and decision-support workflows, model selection has emerged as an important practical challenge. Prior works (e.g., \citealt{brimont2026survey}) primarily evaluates VLMs through benchmark rankings derived from retrieval, captioning, and reasoning downstream tasks using metrics such as Recall@$K$, BLEU, CIDEr, SPICE, and task accuracy. Let $P(M)$ denote the benchmark capability of a model $M$. Under evaluation protocols, models are ranked according to $P(M)$, implicitly assuming that higher capability corresponds to higher reliability. However, models with similar values of $P(M)$ often show substantially different performance distributions across datasets and evaluation conditions. As a result, benchmark rankings may provide limited insight into how consistently a model behaves beyond the evaluation benchmark. We refer to this discrepancy as the \textit{Capability-Reliability Gap}. Let $R(M)$ denote the reliability of a model under varying evaluation conditions. In practice, the assumption $P(M_i) > P(M_j) \Rightarrow R(M_i) > R(M_j)$ frequently fails, implying that the model maximizing capability, $\arg\max_M P(M)$, may not coincide with the model maximizing reliability, $\arg\max_M R(M)$ (see Figure~\ref{fig:capability_reliability_gap}).

\begin{figure*}[t!]
\vspace{-0.3cm}
\centering
    \includegraphics[width=0.85\linewidth]{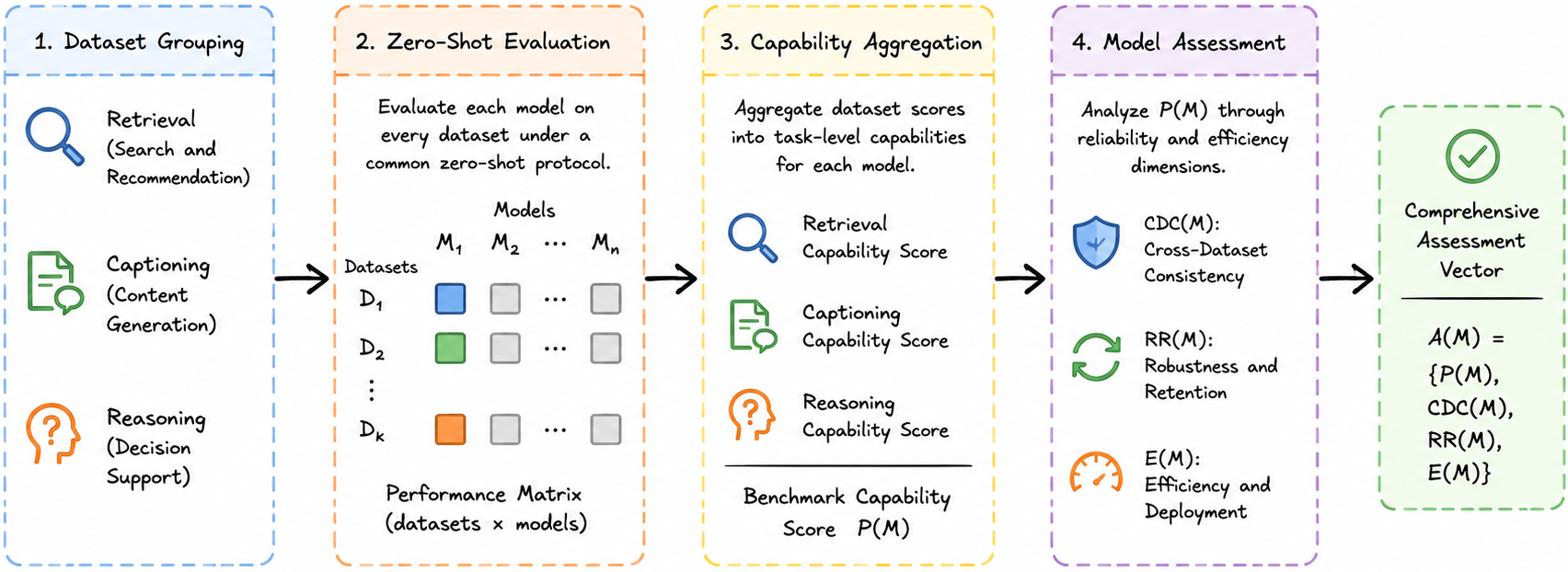}
    \vspace{-0.3cm}
    \caption{Overview of ARGUS-EVAL. Dataset-level results from retrieval, captioning, and reasoning downstream tasks are aggregated into $P(M)$ and subsequently analyzed through $CDC(M)$, $RR(M)$, and $E(M)$ for capability-reliability-oriented VLM assessment.}
    \label{arch}
    \vspace{-0.3cm}
\end{figure*}

To study this gap, we introduce ARGUS-EVAL, a capability-reliability-oriented evaluation framework for VLMs. Unlike traditional evaluations that focus primarily on Benchmark Capability $P(M)$, ARGUS-EVAL extends benchmark assessment through Cross-Dataset Consistency ($CDC(M)$), which measures performance stability across heterogeneous benchmarks, and Robustness Retention ($RR(M)$), which measures performance preservation under image and text perturbations. Along with Efficiency $E(M)$, these measures characterize model behavior through the assessment vector $\mathcal{A}(M)=\{P(M),CDC(M),RR(M),E(M)\}$. Using ARGUS-EVAL, we evaluate CLIP\footnote{\scriptsize{\url{https://huggingface.co/docs/transformers/model_doc/clip}}}, BLIP\footnote{\scriptsize{\url{https://huggingface.co/docs/transformers/model_doc/blip}}}, LXMERT\footnote{\scriptsize{\url{https://huggingface.co/docs/transformers/model_doc/lxmert}}}, Gemma-3-4B\footnote{\scriptsize{\url{https://huggingface.co/google/gemma-3-4b-it}}}, and Qwen-2.5VL-3B-Instruct\footnote{\scriptsize{\url{https://huggingface.co/Qwen/Qwen2.5-VL-3B-Instruct}}} across retrieval, captioning, and reasoning downstream tasks. In summary, the main contributions of this work are as follows:

\begin{itemize}
\vspace{-0.3cm}
    \item We introduce ARGUS-EVAL, a capability-reliability-oriented evaluation framework for VLMs that extends traditional benchmark assessment through $CDC(M)$, $RR(M)$, and $E(M)$.
\vspace{-0.3cm}
    \item We conduct a comprehensive empirical evaluation of CLIP, BLIP, LXMERT, Gemma-3-4B, and Qwen-2.5VL-3B-Instruct across retrieval, captioning, and reasoning tasks.
\end{itemize}

\section{Related Work}

Prior works studies VLMs across retrieval, captioning, and reasoning downstream tasks, as discussed in \S\ref{Introduction}. Works include MMBench \cite{liu2024mmbench}, MMVet \cite{yu2023mm}, MMMU \cite{yue2024mmmu}, SEED-Bench \cite{li2023seed}, MME \cite{fu2026mme}, and Video-MME \cite{fu2025video}, which are widely used to compare models such as GPT-4V \cite{yang2023dawn}, Gemini \cite{comanici2025gemini}, Claude 3 \cite{TheC3}, InternVL \cite{chen2024internvl}, Molmo \cite{deitke2025molmo}, Gemma-3 \cite{Kamath2025Gemma3T}, and Qwen2.5-VL \cite{bai2025qwen3} through benchmark-based rankings. Other works including MM-SafetyBench \cite{liu2024mm}, RobustBench \cite{croce2020robustbench}, AdvCLIP \cite{zhou2023advclip}, and perturbation-based evaluations of CLIP \cite{radford2021learning} and BLIP \cite{li2022blip} studies image corruption, adversarial attacks, OCR degradation, and distribution shifts, while deployment-oriented works studies latency \cite{faraz2026designing}, memory footprint \cite{haddadresource}, quantization \cite{jacob2018quantization}, and inference efficiency \cite{dritsas2026deployment} for multimodal systems \cite{zheng2026multimodal}. Nevertheless, prior works typically assess capability, robustness, or efficiency independently. To the best of our knowledge, prior works do not provide a unified framework for jointly assessing $P(M)$, $CDC(M)$, $RR(M)$, and $E(M)$. ARGUS-EVAL addresses this limitation through a capability-reliability-oriented evaluation framework for VLMs (see Table~\ref{tab:argus_eval_comparison}). 

\begin{table}[t]
\vspace{-0.7cm}
\centering
\scriptsize
\renewcommand{\arraystretch}{1.05}
\setlength{\tabcolsep}{4pt}

\begin{tabular}{lcccc}
\toprule
\textbf{Framework} &
\cellcolor{coral!20}\textbf{$P(M)$} &
\cellcolor{MintCream!195}\textbf{$CDC(M)$} &
\cellcolor{Beige!85}\textbf{$RR(M)$} &
\cellcolor{softblue!25}\textbf{$E(M)$} \\
\midrule
MMBench        & \checkmark & \cross & \cross & \cross \\
MMVet          & \checkmark & \cross & \cross & \cross \\
MMMU           & \checkmark & \cross & \cross & \cross \\
SEED-Bench     & \checkmark & \cross & \cross & \cross \\
MME            & \checkmark & \cross & \cross & \cross \\
Video-MME      & \checkmark & \cross & \cross & \cross \\
MM-SafetyBench & \cross & \cross & \checkmark & \cross \\
MM-RobustBench & \cross & \cross & \checkmark & \cross \\
AdvCLIP        & \checkmark & \cross & \checkmark & \cross \\
\midrule
\textbf{ARGUS-EVAL} &
\cellcolor{coral!20}\checkmark &
\cellcolor{MintCream!195}\checkmark &
\cellcolor{Beige!85}\checkmark &
\cellcolor{softblue!25}\checkmark \\
\bottomrule
\end{tabular}

\vspace{-0.3cm}
\caption{Comparison of SOTAs for industry VLM applications, highlighting that ARGUS-EVAL is the only framework combining all assessment.} \label{tab:argus_eval_comparison}
\vspace{-0.4cm}
\end{table}

\section{Methodology}
\label{Methodology}

ARGUS-EVAL is a capability-reliability-oriented evaluation framework as discussed in \S\ref{Introduction} (see Figure~\ref{arch}). Let $\mathcal{M}=\{M_1,\ldots,M_n\}$ denote a set of VLMs and let $\mathcal{D}=\{D_1,\ldots,D_k\}$ denote a collection of datasets spanning retrieval, captioning, and reasoning downstream tasks. Each dataset $D_i$ consists of multimodal instances $x=(v,t)$, where $v$ and $t$ denote visual and textual inputs respectively. Given a model $M_j$ and dataset instance $x_i\in D_i$, inference is represented as $\hat{y}_{i,j}=f_j(x_i)$. For retrieval tasks, $\hat{y}_{i,j}$ denotes a ranked set of image-text pairs; for captioning tasks, $\hat{y}_{i,j}$ denotes a generated text sequence; and for reasoning tasks, $\hat{y}_{i,j}$ denotes a predicted answer. These task-specific outputs are converted into dataset scores according to the evaluation protocol of the corresponding benchmark.

\begin{figure}[t!]
\vspace{-0.3cm}
\centering
\definecolor{retrievalblue}{RGB}{30,144,255}
\definecolor{captiongreen}{RGB}{46,139,87}
\definecolor{reasoningorange}{RGB}{210,105,30}
\tcbset{
  boxrule=0.2pt,
  arc=2pt,
  left=1.5pt,
  right=1.5pt,
  top=1pt,
  bottom=1pt,
  boxsep=1.5pt,
  before skip=4pt,
  after skip=4pt,
  width=0.9\linewidth
}
\scriptsize

% ===== Retrieval =====
\begin{tcolorbox}[colback=gray!05, colframe=retrievalblue]
\textbf{\textcolor{retrievalblue}{Workflow: Retrieval (\texttt{Search and Recommendation})}}
\end{tcolorbox}
\begin{tcolorbox}[colback=gray!05, colframe=black]
\textbf{Prompt Template:}
\textit{Retrieve the most relevant text description for the given image.}
\end{tcolorbox}
\begin{tcolorbox}[colback=gray!05, colframe=black]
\textbf{Output Example:}
A person riding a red bicycle on a city street.
\end{tcolorbox}

% ===== Captioning =====
\begin{tcolorbox}[colback=gray!05, colframe=captiongreen]
\textbf{\textcolor{captiongreen}{Workflow: Captioning (\texttt{Content Generation})}}
\end{tcolorbox}
\begin{tcolorbox}[colback=gray!05, colframe=black]
\textbf{Prompt Template:}
\textit{Describe the image in one factual sentence.}
\end{tcolorbox}
\begin{tcolorbox}[colback=gray!05, colframe=black]
\textbf{Output Example:}
A group of office workers collaborating around a table.
\end{tcolorbox}

% ===== Reasoning =====
\begin{tcolorbox}[colback=gray!05, colframe=reasoningorange]
\textbf{\textcolor{reasoningorange}{Workflow: Reasoning (\texttt{Decision Support})}}
\end{tcolorbox}
\begin{tcolorbox}[colback=gray!05, colframe=black]
\textbf{Prompt Template:}
\textit{Select the correct answer and provide a brief rationale.}
\end{tcolorbox}
\begin{tcolorbox}[colback=gray!05, colframe=black]
\textbf{Output Example:}
The blue cube is left of the red sphere.\\
Rationale: The image shows the blue cube positioned left of the red sphere.
\end{tcolorbox}
\vspace{-0.1cm}
\caption{
Task-consistent prompt templates used under the common zero-shot evaluation protocol.
}
\label{temp}
\vspace{-0.7cm}
\end{figure}

As illustrated in Figure~\ref{arch}, ARGUS-EVAL follows a four-stage workflow. In the first stage, datasets are grouped according to retrieval, captioning, and reasoning task categories. In the second stage, every model $M_j \in \mathcal{M}$ is evaluated on every dataset $D_i \in \mathcal{D}$ using official pretrained checkpoints under a common zero-shot protocol, producing a performance matrix $\mathbf{S}\in\mathbb{R}^{k\times n}$, where each entry $s_{i,j}$ denotes the performance of model $M_j$ on dataset $D_i$. For instruction-tuned models, task-consistent prompts are used to obtain retrieval, captioning, and reasoning outputs, whereas retrieval-based models operate through their native image-text matching mechanisms (see Figure~\ref{temp}). In the third stage, dataset-level scores are aggregated into retrieval, captioning, and reasoning assessments to obtain benchmark capability $P(M)$. For a model $M_j$, performance across benchmarks is represented by the score vector $\mathbf{s}_j=\{s_{1,j},s_{2,j},\ldots,s_{k,j}\}$, where $s_{i,j}$ denotes the performance of $M_j$ on dataset $D_i$. Traditional benchmark rankings are derived from the aggregate capability score $P(M_j)$ computed from $\mathbf{s}_j$. However, models with similar values of $P(M)$ may show substantially different score distributions across datasets. Therefore, ARGUS-EVAL analyzes the dataset-wise variation of $\mathbf{s}_j$ and its response to controlled image and text perturbations rather than relying solely on the aggregate capability score. In the fourth stage, each model is characterized through the assessment vector $\mathcal{A}(M)=\{P(M),CDC(M),RR(M),E(M)\}$, where $CDC(M)$ quantifies performance stability across heterogeneous benchmarks, $RR(M)$ quantifies performance retention under perturbations, and $E(M)$ captures deployment characteristics including latency, memory footprint, and throughput. The mathematical formulation of these measures is presented in Section~\ref{Evaluation Metrics}.

\section{Experimental Setup}

\subsection{Datasets}

To evaluate VLMs across retrieval, captioning, and reasoning downstream tasks, ARGUS-EVAL uses five widely adopted datasets. Retrieval and captioning are evaluated on COCO\footnote{\scriptsize\url{https://cocodataset.org/\#home}}~\cite{lin2014microsoft}, Flickr30k\footnote{\scriptsize\url{https://forms.illinois.edu/sec/229675}}~\cite{young2014image}, and Visual Genome\footnote{\scriptsize\url{https://homes.cs.washington.edu/~ranjay/visualgenome/api.html}}~\cite{krishna2017visual}, while reasoning is evaluated on CLEVR\footnote{\scriptsize\url{https://cs.stanford.edu/people/jcjohns/clevr/}}~\cite{johnson2017clevr} and VCR\footnote{\scriptsize\url{https://visualcommonsense.com/}}~\cite{zellers2019recognition}. These datasets constitute the dataset collection $\mathcal{D}$ used throughout ARGUS-EVAL. Evaluations use the standard test splits of COCO (5,000 samples), Flickr30k (1,000 samples), Visual Genome (5,000 samples), CLEVR (15,000 samples), and VCR (25,263 samples).

\subsection{Evaluation Metrics}
\label{Evaluation Metrics}

ARGUS-EVAL characterizes each model through the assessment vector $\mathcal{A}(M)=\{P(M),CDC(M),RR(M),E(M)\}$. The definitions of these measures are presented below, and their theoretical properties are provided in Appendix~\ref{Theoretical Properties}.

\vspace{-0.3cm}
\paragraph{Benchmark Capability $P(M)$.}
Benchmark capability represents the aggregate performance of a model across retrieval, captioning, and reasoning downstream tasks. Let $\mathbf{s}_j=\{s_{1,j},s_{2,j},\ldots,s_{k,j}\}$ denote the normalized benchmark score vector of model $M_j$, where $s_{i,j}\in[0,1]$ denotes the performance of $M_j$ on dataset $D_i$. Benchmark capability is computed as $P(M_j)=\frac{1}{k}\sum_{i=1}^{k}s_{i,j}$, where $P(M_j)\in[0,1]$ and higher values indicate stronger overall benchmark performance.

\vspace{-0.3cm}
\paragraph{Retrieval Tasks.}
For retrieval evaluation on COCO, Flickr30k, and Visual Genome, we report Recall@$K$ ($R@K$), Mean Reciprocal Rank (MRR), and Median Rank (MedR). Recall@$K$ for $K\in\{1,5,10\}$ measures the proportion of queries whose relevant item $y_i^\ast$ appears within the top-$K$ retrieved results: $R@K=\frac{1}{N}\sum_{i=1}^{N}\mathbb{1}[\mathrm{rank}(y_i^\ast)\leq K]$. Mean Reciprocal Rank is defined as $\mathrm{MRR}=\frac{1}{N}\sum_{i=1}^{N}\frac{1}{\mathrm{rank}(y_i^\ast)}$, while Median Rank is computed as $\mathrm{MedR}=\mathrm{median}_{i}(\mathrm{rank}(y_i^\ast))$. Higher values indicate better performance for $R@K$ and MRR, whereas lower values indicate better performance for MedR. COCO and Flickr30k use Image$\rightarrow$Text retrieval, while Visual Genome uses Text$\rightarrow$Image retrieval. All retrieval metrics are normalized to $[0,1]$ before computing $P(M)$.

\vspace{-0.3cm}
\paragraph{Captioning Tasks.}
For caption generation on COCO and Flickr30k, we report BLEU, METEOR, CIDEr, and SPICE. BLEU evaluates $n$-gram precision through $\mathrm{BLEU}_n=\exp\!\left(\sum_{k=1}^{n}w_k\log p_k\right)$, where $p_k$ denotes matched $k$-gram precision. METEOR is defined as $\mathrm{METEOR}=F_{\mathrm{mean}}(1-P_{\mathrm{penalty}})$, where $F_{\mathrm{mean}}=\frac{10PR}{R+9P}$. CIDEr measures agreement with reference captions: $\mathrm{CIDEr}=\frac{1}{N}\sum_{i=1}^{N}\frac{\hat{C}_i\cdot C_i}{\|\hat{C}_i\|\|C_i\|}$. SPICE evaluates semantic agreement through scene-graph matching: $\mathrm{SPICE}=F_1(\mathrm{scene\mbox{-}graph}(\hat{C}_i),\mathrm{scene\mbox{-}graph}(C_i))$. Higher values indicate better performance for all metrics. All captioning metrics are normalized to $[0,1]$ before computing $P(M)$.

\begin{table*}[t!]
\vspace{-0.3cm}
\centering
\scriptsize
\setlength{\tabcolsep}{3pt}
\renewcommand{\arraystretch}{1.05}

\begin{tabular}{l|
ccccc|
ccccc|
ccccc}
\toprule

\textbf{Model} &
\multicolumn{5}{c|}{\textbf{COCO (Image $\rightarrow$ Text)}} &
\multicolumn{5}{c|}{\textbf{Flickr30k (Image $\rightarrow$ Text)}} &
\multicolumn{5}{c}{\textbf{Visual Genome (Text $\rightarrow$ Image)}} \\

& \cellcolor{PeachPuff!70}R@1
& \cellcolor{LightCyan!70}R@5
& \cellcolor{LemonChiffon!80}R@10
& \cellcolor{Lavender!80}MRR
& \cellcolor{LightPink!80}MedR

& \cellcolor{PeachPuff!70}R@1
& \cellcolor{LightCyan!70}R@5
& \cellcolor{LemonChiffon!80}R@10
& \cellcolor{Lavender!80}MRR
& \cellcolor{LightPink!80}MedR

& \cellcolor{PeachPuff!70}R@1
& \cellcolor{LightCyan!70}R@5
& \cellcolor{LemonChiffon!80}R@10
& \cellcolor{Lavender!80}MRR
& \cellcolor{LightPink!80}MedR \\
\midrule

LXMERT
& 71.2 & 89.5 & 94.8 & 0.77 & 2
& 68.9 & 88.0 & 92.5 & 0.74 & 3
& 70.4 & 89.6 & 94.2 & 0.76 & 2 \\

CLIP
& 74.0 & 91.3 & 95.7 & 0.80 & 2
& 72.1 & 89.5 & 94.1 & 0.78 & 2
& 66.8 & 84.5 & 91.1 & 0.72 & 3 \\

BLIP
& 78.9 & 94.2 & 97.6 & 0.84 & 1
& 76.5 & 93.3 & 96.8 & 0.82 & 1
& 68.3 & 87.1 & 93.9 & 0.75 & 2 \\

Gemma
& 80.2 & 95.1 & 97.9 & 0.86 & 1
& 78.3 & 94.1 & 97.2 & 0.84 & 1
& 71.5 & 89.8 & 94.6 & 0.79 & 1 \\

Qwen

& \cellcolor{PeachPuff!70}\textbf{82.7}
& \cellcolor{LightCyan!70}\textbf{96.4}
& \cellcolor{LemonChiffon!80}\textbf{98.3}
& \cellcolor{Lavender!80}\textbf{0.88}
& \cellcolor{LightPink!80}\textbf{1}

& \cellcolor{PeachPuff!70}\textbf{80.5}
& \cellcolor{LightCyan!70}\textbf{95.7}
& \cellcolor{LemonChiffon!80}\textbf{97.9}
& \cellcolor{Lavender!80}\textbf{0.87}
& \cellcolor{LightPink!80}\textbf{1}

& \cellcolor{PeachPuff!70}\textbf{73.2}
& \cellcolor{LightCyan!70}\textbf{90.6}
& \cellcolor{LemonChiffon!80}\textbf{95.4}
& \cellcolor{Lavender!80}\textbf{0.81}
& \cellcolor{LightPink!80}\textbf{1} \\

\bottomrule
\end{tabular}

\vspace{-0.3cm}
\caption{
Cross-dataset image-text and text-image retrieval performance on COCO, Flickr30k, and Visual Genome. Higher R@K and MRR indicate better retrieval performance, while lower MedR indicates better ranking quality.
}
\label{tab:retrieval-results}
\vspace{-0.4cm}
\end{table*}

\begin{table*}[t!]
%\vspace{-0.1cm}
\centering
\scriptsize
\setlength{\tabcolsep}{2pt}
\renewcommand{\arraystretch}{1.05}
\begin{tabular*}{\textwidth}{@{\extracolsep{\fill}}lccccccc|ccccccc@{}}
\toprule
\textbf{Model}
& \multicolumn{7}{c|}{\textbf{COCO (Captioning)}}
& \multicolumn{7}{c}{\textbf{Flickr30k (Captioning)}} \\
\cmidrule(lr){2-8}
\cmidrule(lr){9-15}

& \cellcolor{PeachPuff!70}BLEU-1
& \cellcolor{LightCyan!70}BLEU-2
& \cellcolor{LemonChiffon!80}BLEU-3
& \cellcolor{Lavender!80}BLEU-4
& \cellcolor{LightPink!80}METEOR
& \cellcolor{MintCream!80}CIDEr
& \cellcolor{Apricot!60}SPICE

& \cellcolor{PeachPuff!70}BLEU-1
& \cellcolor{LightCyan!70}BLEU-2
& \cellcolor{LemonChiffon!80}BLEU-3
& \cellcolor{Lavender!80}BLEU-4
& \cellcolor{LightPink!80}METEOR
& \cellcolor{MintCream!80}CIDEr
& \cellcolor{Apricot!60}SPICE \\
\midrule

LXMERT
& 75.3 & 62.7 & 48.5 & 34.1 & 25.4 & 90.3 & 18.9
& 70.8 & 59.5 & 44.7 & 31.5 & 23.8 & 75.1 & 17.1 \\

CLIP
& 62.1 & 50.9 & 38.7 & 22.3 & 19.2 & 65.7 & 14.2
& 59.5 & 48.7 & 36.4 & 19.7 & 16.3 & 55.4 & 12.5 \\

BLIP
& 78.9 & 66.4 & 50.8 & 43.2 & 29.7 & 134.5 & 22.4
& 74.5 & 61.9 & 46.2 & 39.7 & 27.3 & 95.2 & 20.8 \\

Gemma
& 80.6 & 68.9 & 53.5 & 45.8 & 31.4 & 138.1 & 23.1
& 76.2 & 63.8 & 48.1 & 41.2 & 28.5 & 98.9 & 21.6 \\

Qwen

& \cellcolor{PeachPuff!70}\textbf{82.4}
& \cellcolor{LightCyan!70}\textbf{70.3}
& \cellcolor{LemonChiffon!80}\textbf{55.1}
& \cellcolor{Lavender!80}\textbf{47.2}
& \cellcolor{LightPink!80}\textbf{32.1}
& \cellcolor{MintCream!80}\textbf{141.6}
& \cellcolor{Apricot!60}\textbf{23.8}

& \cellcolor{PeachPuff!70}\textbf{78.1}
& \cellcolor{LightCyan!70}\textbf{65.4}
& \cellcolor{LemonChiffon!80}\textbf{49.3}
& \cellcolor{Lavender!80}\textbf{42.7}
& \cellcolor{LightPink!80}\textbf{29.1}
& \cellcolor{MintCream!80}\textbf{101.3}
& \cellcolor{Apricot!60}\textbf{22.3} \\
\bottomrule
\end{tabular*}

\vspace{-0.3cm}
\caption{
Zero-shot captioning performance on COCO and Flickr30k. Higher BLEU, METEOR, CIDEr, and SPICE indicate better caption quality.
}
\label{tab:caption-results}
\vspace{-0.4cm}
\end{table*}

\vspace{-0.3cm}
\paragraph{Reasoning Tasks.}
Reasoning performance is evaluated on CLEVR and VCR. For CLEVR, we report classification accuracy: $\mathrm{Accuracy}=\frac{1}{N}\sum_{i=1}^{N}\mathbb{1}[a_i=a_i^\ast]$. For VCR, we follow the standard evaluation protocol and report Question-to-Answer (Q$\rightarrow$A) and Question-Answer-to-Rationale (QA$\rightarrow$R) accuracy. Higher values indicate better performance for all reasoning metrics. All reasoning metrics are normalized to $[0,1]$ before computing $P(M)$.

\vspace{-0.3cm}
\paragraph{Cross-Dataset Consistency ($CDC$).}
CDC measures performance stability across heterogeneous benchmarks. Let $\bar{s}_j=\frac{1}{k}\sum_{i=1}^{k}s_{i,j}$ denote the mean normalized benchmark score of model $M_j$. Then, $CDC(M_j)=1-\frac{1}{k}\sum_{i=1}^{k}\left|\frac{s_{i,j}-\bar{s}_j}{\bar{s}_j}\right|$, where higher values indicate smaller performance variation across datasets.

\vspace{-0.3cm}
\paragraph{Robustness Retention ($RR$).}
RR measures performance preservation under image and text perturbations. Let $P_{\mathrm{clean}}(M)$ and $P_{\mathrm{perturbed}}(M)$ denote benchmark capability before and after perturbation. Then, $RR(M)=\frac{P_{\mathrm{perturbed}}(M)}{P_{\mathrm{clean}}(M)}$, where higher values indicate stronger robustness.

\vspace{-0.3cm}
\paragraph{Efficiency $E(M)$.}
Efficiency is measured through inference latency, memory footprint, FLOPs, and throughput. Latency measures average inference time per sample: $\mathrm{Latency}=\frac{1}{N}\sum_{i=1}^{N}t_i$. Memory footprint is computed as $\mathrm{Memory}=\frac{\mathrm{Mem}_{\mathrm{peak}}}{B}$, where $\mathrm{Mem}_{\mathrm{peak}}$ denotes peak GPU memory allocation for batch size $B$. FLOPs denote the total floating-point operations per sample. Throughput measures samples processed per second: $\mathrm{Throughput}=\frac{N}{\sum_{i=1}^{N}t_i}$. Lower values indicate better performance for Latency, Memory, and FLOPs, whereas higher values indicate better performance for Throughput; all efficiency metrics are normalized to $[0,1]$ before computing $E(M)$. Let $\mathbf{e}_j=\{e_{1,j},e_{2,j},e_{3,j},e_{4,j}\}$ denote the normalized latency, memory, FLOPs, and throughput scores of model $M_j$. Then, $E(M_j)=\frac{1}{4}\sum_{r=1}^{4}e_{r,j}$.

\subsection{Hyperparameters}

All experiments use official pretrained checkpoints under a zero-shot protocol without fine-tuning. Inference is performed on NVIDIA A100 GPUs (40\,GB VRAM) with batch size $32$ and FP16 precision. Images are resized to $224\times224$ pixels. Retrieval evaluation uses Recall@$K$ with $K\in\{1,5,10\}$, while generative models use greedy decoding (temperature $=0$) with maximum output length $40$ tokens. Results are averaged over $3$ random seeds.

\subsection{Baselines}

We evaluate five representative VLMs: CLIP, BLIP, LXMERT, Gemma-3-4B (Gemma), and Qwen-2.5VL-3B-Instruct (Qwen). CLIP is a contrastive image-text model for retrieval, BLIP combines vision-language pretraining with text generation, and LXMERT is a cross-modal transformer for vision-language understanding. Gemma-3-4B and Qwen-2.5VL-3B-Instruct are instruction-tuned VLMs evaluated across retrieval, captioning, and downstream reasoning tasks. All models are evaluated using official pretrained checkpoints under the common zero-shot protocol described in Section~\ref{Methodology}.

\section{Downstream Analysis}
\label{Downstream Analysis}

\paragraph{Retrieval Task.}
Table~\ref{tab:retrieval-results} reports retrieval performance across COCO, Flickr30k, and Visual Genome. Qwen-2.5VL-3B-Instruct achieves the strongest overall retrieval performance, obtaining the highest R@1, R@5, R@10, and MRR across all datasets, with R@1 scores of 82.7\%, 80.5\%, and 73.2\% on COCO, Flickr30k, and Visual Genome, respectively. Gemma-3-4B consistently ranks second, while BLIP generally outperforms CLIP and LXMERT. These results indicate that the instruction-tuned VLMs evaluated in this study achieve stronger image-text retrieval performance than the retrieval-oriented baselines, suggesting their suitability for industry VLM applications such as search and recommendation systems where accurate top-ranked retrieval is important.

\vspace{-0.3cm}
\paragraph{Caption Task.}
Table~\ref{tab:caption-results} reports zero-shot captioning performance on COCO and Flickr30k. Qwen-2.5VL-3B-Instruct achieves the strongest overall results, obtaining the highest BLEU-4, METEOR, CIDEr, and SPICE scores on both datasets, including BLEU-4 scores of 47.2 and 42.7 and CIDEr scores of 141.6 and 101.3 on COCO and Flickr30k, respectively. Gemma-3-4B consistently ranks second, followed by BLIP and LXMERT, while CLIP records the lowest scores across most metrics. The results indicate that the instruction-tuned VLMs evaluated in this study produce more accurate and semantically aligned captions, supporting their suitability for industry content-generation applications where caption quality is important.

\begin{table}[t!]
\vspace{-0.6cm}
\centering
\scriptsize
\begin{tabular}{l|c|c|c}
\toprule
\textbf{Model} &
\cellcolor{PeachPuff!70}\textbf{CLEVR} &
\cellcolor{LightCyan!70}\textbf{VCR Q$\rightarrow$A} &
\cellcolor{Lavender!80}\textbf{VCR QA$\rightarrow$R} \\
\midrule

LXMERT      & 96.3 & 70.2 & 71.5 \\
CLIP        & 84.5 & 62.3 & 58.9 \\
BLIP        & 89.7 & 74.8 & 68.4 \\
Gemma-3-4B  & 94.1 & 77.2 & 73.9 \\
Qwen-2.5VL-3B-Instruct
& \cellcolor{PeachPuff!70}\textbf{97.4}
& \cellcolor{LightCyan!70}\textbf{79.5}
& \cellcolor{Lavender!80}\textbf{75.8} \\

\bottomrule
\end{tabular}

\vspace{-0.3cm}
\caption{
Zero-shot reasoning performance on CLEVR and VCR. Higher values indicate better reasoning accuracy.
}
\label{tab:reasoning}
\vspace{-0.4cm}
\end{table}

\begin{figure}[t]
\vspace{-0.1cm}
    \centering
    \includegraphics[width=\linewidth,height=6cm,keepaspectratio]{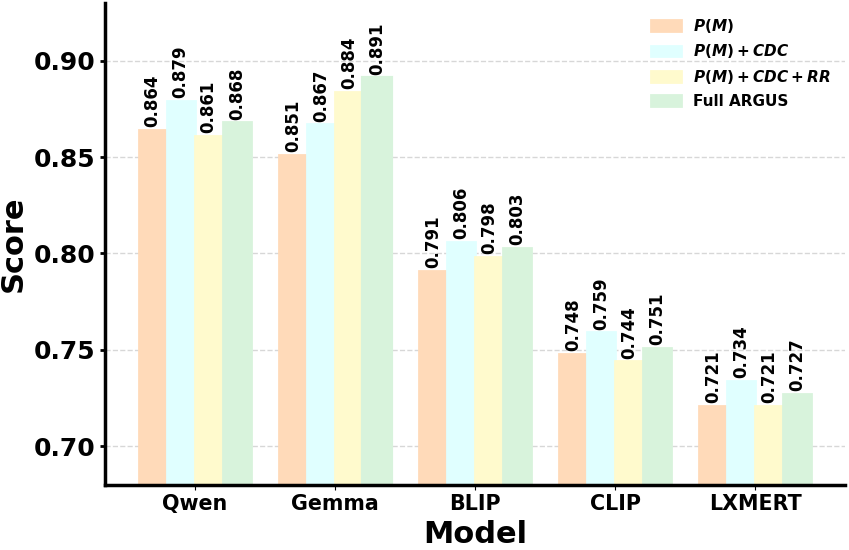}
    \vspace{-0.7cm}
    \caption{
Ablation analysis showing the effect of progressively incorporating reliability-oriented assessment components.
}
    \label{tab:ranking_ablation}
    \vspace{-0.4cm}
\end{figure}

\begin{comment}
    
\begin{table}[t!]
\vspace{-0.1cm}
\centering
\scriptsize
\renewcommand{\arraystretch}{1.05}
\setlength{\tabcolsep}{5pt}

\begin{tabular}{l|c|c|c|c|c}
\toprule
\textbf{Assessment} &
\cellcolor{PeachPuff!70}\textbf{Qwen} &
\cellcolor{LightCyan!70}\textbf{Gemma} &
\cellcolor{LemonChiffon!80}\textbf{BLIP} &
\cellcolor{Lavender!80}\textbf{CLIP} &
\cellcolor{MintCream!80}\textbf{LXMERT} \\
\midrule

\textbf{$P(M)$}
& \cellcolor{PeachPuff!70}\textbf{0.864}
& 0.851
& 0.791
& 0.748
& 0.721 \\

\textbf{$P(M)+CDC$}
& \cellcolor{PeachPuff!70}\textbf{0.879}
& 0.867
& 0.806
& 0.759
& 0.734 \\

\textbf{$P(M)+CDC+RR$}
& 0.861
& \cellcolor{LightCyan!70}\textbf{0.884}
& 0.798
& 0.744
& 0.721 \\

\textbf{Full ARGUS}
& 0.868
& \cellcolor{LightCyan!70}\textbf{0.891}
& 0.803
& 0.751
& 0.727 \\

\bottomrule
\end{tabular}

\vspace{-0.3cm}
\caption{
Ablation analysis showing the effect of progressively incorporating reliability-oriented assessment components.
}
\label{tab:ranking_ablation}
\vspace{-0.4cm}
\end{table}

\end{comment}

\begin{figure*}[t!]
    \centering

    \begin{subfigure}{0.32\textwidth}
        \centering
        \includegraphics[width=\linewidth]{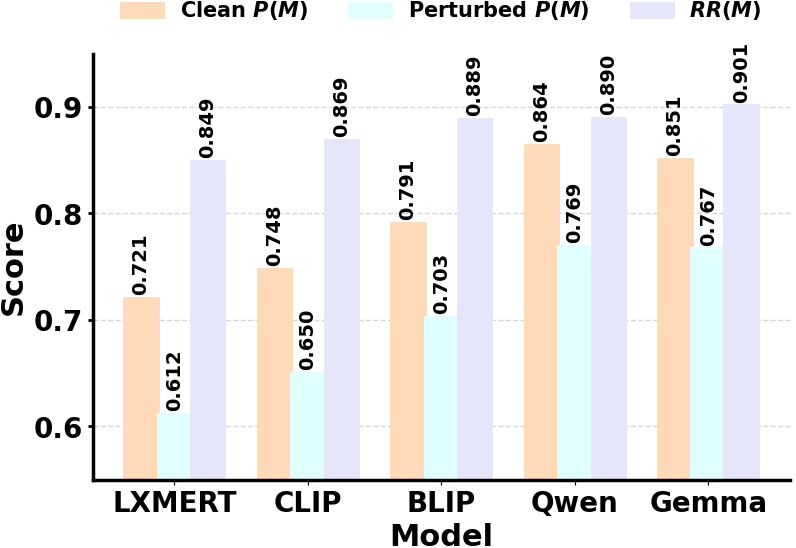}
        \caption{Robustness Retention}
    \end{subfigure}%
    \begin{subfigure}{0.32\textwidth}
        \centering
        \includegraphics[width=\linewidth]{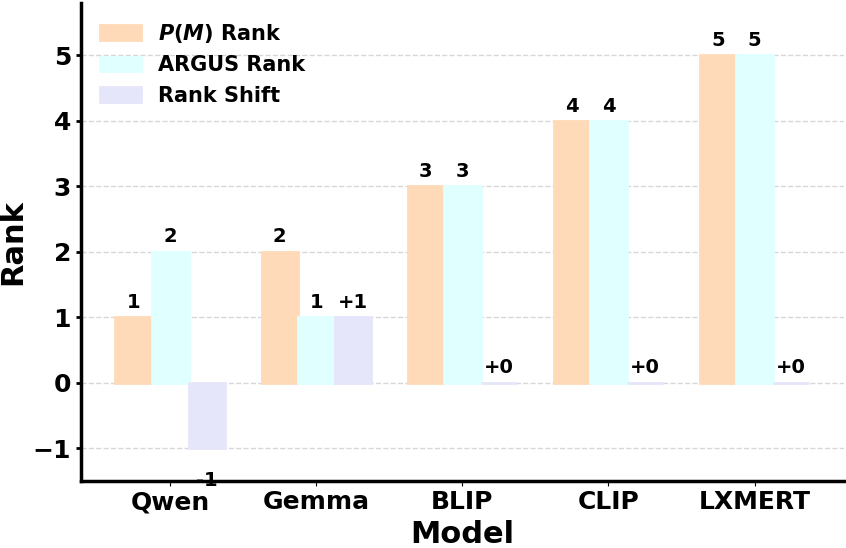}
        \caption{Capability-Reliability Gap}
    \end{subfigure}%
    \begin{subfigure}{0.32\textwidth}
        \centering
        \includegraphics[width=\linewidth]{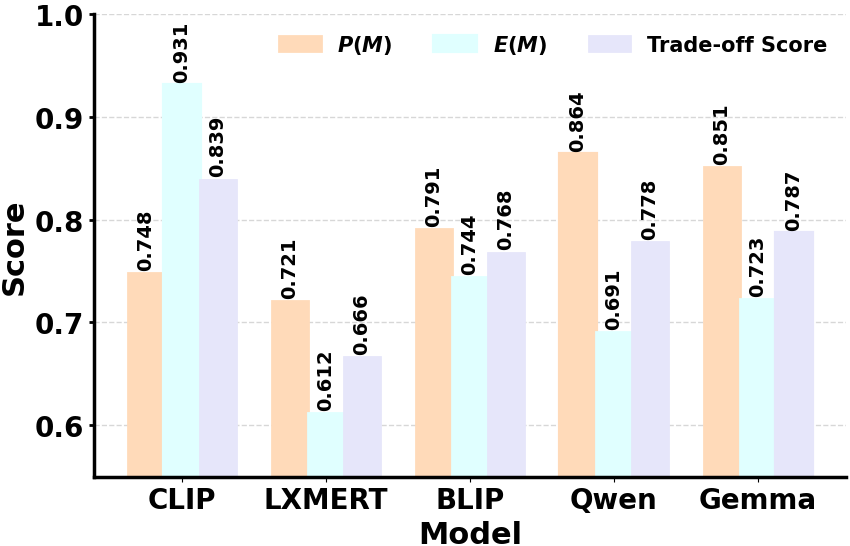}
        \caption{Efficiency-Capability Trade-off}
    \end{subfigure}

    \vspace{-0.3cm}
    \caption{
Empirical analysis of ARGUS-EVAL assessment dimensions. (a) Robustness Retention ($RR$) under image and text perturbations, showing stronger retention for Gemma under distribution shifts. (b) Capability-Reliability Gap, where incorporating reliability measures reverses the top ranking between Qwen and Gemma. (c) Efficiency-Capability Trade-off, highlighting the relationship between benchmark capability and deployment efficiency. 
}
    \label{fig:argus_analysis}
\vspace{-0.3cm}
\end{figure*}

\begin{table*}[t!]
\vspace{-0.1cm}
\centering
\scriptsize
\setlength{\tabcolsep}{2pt}
\renewcommand{\arraystretch}{1.05}

\begin{tabular*}{\textwidth}{@{\extracolsep{\fill}}lcccc|cccc|cccc@{}}
\toprule
\multirow{2}{*}{\textbf{Model}}
& \multicolumn{4}{c|}{\textbf{NVIDIA A100 (40GB)}} 
& \multicolumn{4}{c|}{\textbf{Jetson Nano (4GB)}} 
& \multicolumn{4}{c}{\textbf{Raspberry Pi 5 (8GB)}} \\
\cmidrule(lr){2-5}
\cmidrule(lr){6-9}
\cmidrule(lr){10-13}

& \cellcolor{PeachPuff!70}Lat.
& \cellcolor{LightCyan!70}MF
& \cellcolor{Lavender!80}FLOPs
& \cellcolor{LemonChiffon!80}Thp.

& \cellcolor{PeachPuff!70}Lat.
& \cellcolor{LightCyan!70}MF
& \cellcolor{Lavender!80}FLOPs
& \cellcolor{LemonChiffon!80}Thp.

& \cellcolor{PeachPuff!70}Lat.
& \cellcolor{LightCyan!70}MF
& \cellcolor{Lavender!80}FLOPs
& \cellcolor{LemonChiffon!80}Thp. \\
\midrule

CLIP
& \cellcolor{PeachPuff!70}\textbf{31}
& \cellcolor{LightCyan!70}\textbf{0.9}
& \cellcolor{Lavender!80}\textbf{36}
& \cellcolor{LemonChiffon!80}\textbf{32}
& \cellcolor{PeachPuff!70}\textbf{210}
& \cellcolor{LightCyan!70}\textbf{1.2}
& \cellcolor{Lavender!80}\textbf{36}
& \cellcolor{LemonChiffon!80}\textbf{4.8}
& \cellcolor{PeachPuff!70}\textbf{395}
& \cellcolor{LightCyan!70}\textbf{1.4}
& \cellcolor{Lavender!80}\textbf{36}
& \cellcolor{LemonChiffon!80}\textbf{2.5} \\

BLIP
& 126 & 3.6 & 113 & 13
& 860 & 3.9 & 113 & 1.2
& 1210 & 4.2 & 113 & 0.8 \\

LXMERT
& 161 & 4.3 & 212 & 9
& 980 & 4.5 & 212 & 0.9
& 1350 & 4.8 & 212 & 0.6 \\

Gemma
& 138 & 3.8 & 128 & 12
& 910 & 4.1 & 128 & 1.1
& 1285 & 4.4 & 128 & 0.7 \\

Qwen
& 142 & 4.0 & 133 & 11
& 940 & 4.3 & 133 & 1.0
& 1320 & 4.6 & 133 & 0.7 \\

\bottomrule
\end{tabular*}

\vspace{-0.3cm}
\caption{
Cross-hardware efficiency comparison on NVIDIA A100, Jetson Nano, and Raspberry Pi 5. Lower Latency (Lat.), Memory Footprint (MF), and FLOPs are better, while higher Throughput (Thp.) is better.
}
\label{tab:efficiency-crosshardware}
\vspace{-0.4cm}
\end{table*}

\vspace{-0.3cm} 
\paragraph{Reasoning Task.}
Table~\ref{tab:reasoning} reports zero-shot reasoning performance on CLEVR and VCR. Qwen-2.5VL-3B-Instruct achieves the strongest overall results, obtaining the highest accuracy on CLEVR (97.4\%), VCR Q$\rightarrow$A (79.5\%), and VCR QA$\rightarrow$R (75.8\%). LXMERT performs competitively on CLEVR (96.3\%), while Gemma-3-4B achieves the second-highest performance on both VCR benchmarks with accuracies of 77.2\% and 73.9\%, respectively. The results suggest that recent instruction-tuned VLMs provide stronger visual and commonsense reasoning capabilities than earlier architectures, supporting their use in industry VLM applications that require multimodal understanding and reasoning over visual content.

\section{Ablation Study}

Figure~\ref{tab:ranking_ablation} presents the effect of progressively incorporating reliability-oriented assessment components into ARGUS-EVAL. Using $P(M)$, Qwen achieves the highest score (0.864), followed by Gemma (0.851). After incorporating $CDC$, both models improve, while Qwen retains the highest score (0.879 versus 0.867). However, the inclusion of $RR$ changes the overall assessment, with Gemma achieving the highest score (0.884) and surpassing Qwen (0.861). The same trend is observed under the full ARGUS-EVAL assessment, where Gemma attains the highest overall score (0.891) compared to Qwen (0.868). These results demonstrate that strong benchmark capability alone does not necessarily correspond to the strongest reliability-aware assessment, highlighting the Capability--Reliability Gap and supporting the need to jointly evaluate capability and reliability in deployment-oriented VLM assessment.

\vspace{-0.4cm}
\paragraph{Additional Analysis.}

Figure~\ref{fig:argus_analysis}(a) shows robustness retention under image and text perturbations, where Gemma achieves the highest score ($RR=0.901$), slightly outperforming Qwen ($RR=0.890$). Figure~\ref{fig:argus_analysis}(b) highlights the Capability-Reliability Gap: although Qwen attains the highest benchmark capability ($P(M)=0.864$), Gemma achieves the highest overall ARGUS-EVAL score (0.891) after incorporating consistency and robustness measures. Figure~\ref{fig:argus_analysis}(c) illustrates the Efficiency-Capability trade-off, with CLIP providing the strongest efficiency profile, Qwen the highest capability, and Gemma the most balanced overall assessment. These findings support the ARGUS-EVAL hypothesis that deployment-oriented VLM evaluation should jointly consider capability, reliability, and efficiency.

\vspace{-0.4cm}
\paragraph{Computational Efficiency.}
Table~\ref{tab:efficiency-crosshardware} reports computational efficiency across server-grade and edge hardware platforms. CLIP achieves the strongest efficiency profile, obtaining the lowest latency, memory footprint, and FLOPs while maintaining the highest throughput on all devices. On the NVIDIA A100, CLIP achieves 31 ms latency and 32 samples/s throughput, compared with 138 ms and 12 samples/s for Gemma and 142 ms and 11 samples/s for Qwen. Similar trends are observed on Jetson Nano and Raspberry Pi 5. Among the instruction-tuned VLMs, Gemma consistently shows slightly better efficiency than Qwen across all hardware platforms. When considered alongside the downstream-task results, these findings reveal a capability-efficiency trade-off: Qwen achieves the strongest retrieval, captioning, and reasoning performance, whereas Gemma provides a more favorable efficiency profile with only a modest reduction in benchmark capability.

\section{Conclusion}
\label{sec:Conclusions}

This work presented ARGUS-EVAL, a capability-reliability-oriented evaluation framework for deployment-oriented VLM assessment. Experimental results show that Qwen achieves the strongest retrieval, captioning, and reasoning downstream performance, whereas Gemma attains the highest reliability-aware assessment after incorporating consistency and robustness measures. %Furthermore, CLIP provides the most efficient deployment profile across edge hardware.  

\section*{Limitations}
\label{sec:Limitations}

ARGUS-EVAL evaluates VLMs across representative retrieval, captioning, and reasoning datasets; however, it does not cover the full diversity of real-world deployment scenarios. The framework focuses on zero-shot evaluation and does not assess fine-tuned or task-specific model adaptations. Furthermore, robustness analysis is limited to a predefined set of image and text perturbations and may not capture all distribution shifts encountered in practice. Efficiency measurements are also conducted on a limited set of hardware platforms and may not generalize to all deployment environments. Future work will extend ARGUS-EVAL to additional tasks, robustness settings, and hardware configurations to further improve deployment-oriented VLM assessment.

\section*{Ethics Statement}
\label{sec:Ethics Statement}

This work evaluates publicly available VLMs using standard benchmark datasets and does not involve human participants, personal data collection, or model training. ARGUS-EVAL is designed to support responsible model assessment rather than relying solely on benchmark performance. The framework is intended as an evaluation methodology and does not endorse or recommend any specific model for deployment. We therefore encourage practitioners to complement ARGUS-EVAL with domain-specific risk and governance evaluations before deployment.

\bibliography{main}

\appendix

\section{Appendix}

\subsection{Theoretical Properties of Evaluation Metrics}
\label{Theoretical Properties}

The following lemmas establish fundamental properties of the assessment measures used in ARGUS-EVAL. These properties provide mathematical interpretation of $P(M)$, $CDC(M)$, $RR(M)$, and $E(M)$.

\begin{lemma}[Bounded Benchmark Capability]
Let $s_{i,j}\in[0,1]$ denote the normalized benchmark score of model $M_j$ on dataset $D_i$. If benchmark capability is defined as $P(M_j)=\frac{1}{k}\sum_{i=1}^{k}s_{i,j}$, then $0\leq P(M_j)\leq 1$.
\end{lemma}

\begin{proof}
Since $s_{i,j}\in[0,1]$ for all $i$, their arithmetic mean also lies in $[0,1]$. Therefore, $0\leq \frac{1}{k}\sum_{i=1}^{k}s_{i,j}\leq 1$, which implies $0\leq P(M_j)\leq 1$.
\end{proof}

\textit{Example.}
Suppose a model achieves normalized benchmark scores $\{0.72,0.81,0.87\}$. Then $P(M)=\frac{0.72+0.81+0.87}{3}=0.80$. Therefore, benchmark capability remains on a common scale between 0 and 1, allowing direct comparison across models.

\begin{lemma}[Maximum Cross-Dataset Consistency]
Let $\bar{s}_j=\frac{1}{k}\sum_{i=1}^{k}s_{i,j}$. If $s_{i,j}=\bar{s}_j$ for all datasets $D_i$, then $CDC(M_j)=1$.
\end{lemma}

\begin{proof}
By definition, $CDC(M_j)=1-\frac{1}{k}\sum_{i=1}^{k}\left|\frac{s_{i,j}-\bar{s}_j}{\bar{s}_j}\right|$. If $s_{i,j}=\bar{s}_j$ for every dataset, then each deviation term becomes zero. Therefore, $CDC(M_j)=1$.
\end{proof}

\textit{Example.}
Suppose a model achieves normalized benchmark scores $\{0.80,0.80,0.80,0.80\}$. Since performance is identical across all datasets, the deviation terms vanish and the model achieves the maximum consistency score $CDC(M)=1$.

\begin{lemma}[Scale Invariance of CDC]
Let $s'_{i,j}=\alpha s_{i,j}$ for some constant $\alpha>0$. Then $CDC(M'_j)=CDC(M_j)$.
\end{lemma}

\begin{proof}
Let $\bar{s}'_j=\alpha\bar{s}_j$. Then $\left|\frac{s'_{i,j}-\bar{s}'_j}{\bar{s}'_j}\right|=\left|\frac{\alpha s_{i,j}-\alpha\bar{s}_j}{\alpha\bar{s}_j}\right|=\left|\frac{s_{i,j}-\bar{s}_j}{\bar{s}_j}\right|$. Therefore, all deviation terms remain unchanged, and  $CDC(M'_j)=CDC(M_j)$.
\end{proof}

\textit{Example.}
Consider benchmark scores $\{0.60,0.70,0.80\}$. Multiplying all scores by 10 produces $\{6.0,7.0,8.0\}$. Although the scale changes, the relative variation remains identical, and therefore the CDC value remains unchanged.

\begin{lemma}[Robustness Preservation]
If benchmark capability remains unchanged under perturbation, that is, $P_{\mathrm{perturbed}}(M)=P_{\mathrm{clean}}(M)$, then $RR(M)=1$.
\end{lemma}

\begin{proof}
From the definition, $RR(M)=\frac{P_{\mathrm{perturbed}}(M)}{P_{\mathrm{clean}}(M)}$. Substituting $P_{\mathrm{perturbed}}(M)=P_{\mathrm{clean}}(M)$ gives $RR(M)=1$.
\end{proof}

\textit{Example.}
Assume a model achieves $P_{\mathrm{clean}}(M)=0.84$ on the original benchmark and $P_{\mathrm{perturbed}}(M)=0.84$ after image blur or text corruption. Then $RR(M)=\frac{0.84}{0.84}=1$, indicating complete retention of benchmark capability.

\begin{lemma}[Performance Degradation Under Perturbation]
If $P_{\mathrm{perturbed}}(M)<P_{\mathrm{clean}}(M)$ and $P_{\mathrm{clean}}(M)>0$, then $RR(M)<1$.
\end{lemma}

\begin{proof}
Since $P_{\mathrm{perturbed}}(M)<P_{\mathrm{clean}}(M)$ and $P_{\mathrm{clean}}(M)>0$, dividing both sides by $P_{\mathrm{clean}}(M)$ yields $\frac{P_{\mathrm{perturbed}}(M)}{P_{\mathrm{clean}}(M)}<1$. Therefore, $RR(M)<1$.
\end{proof}

\textit{Example.}
Suppose a model achieves $P_{\mathrm{clean}}(M)=0.90$ and $P_{\mathrm{perturbed}}(M)=0.72$. Then $RR(M)=\frac{0.72}{0.90}=0.80$, indicating that the model retains 80\% of its original benchmark capability after perturbation.

\begin{lemma}[Bounded Efficiency]
Let $\mathbf{e}_j=\{e_{1,j},e_{2,j},e_{3,j},e_{4,j}\}$ denote normalized efficiency scores corresponding to latency, memory footprint, FLOPs, and throughput, where $e_{r,j}\in[0,1]$. If efficiency is defined as $E(M_j)=\frac{1}{4}\sum_{r=1}^{4}e_{r,j}$, then $0\leq E(M_j)\leq 1$.
\end{lemma}

\begin{proof}
Since each normalized efficiency score satisfies $e_{r,j}\in[0,1]$, their arithmetic mean must also lie in $[0,1]$. Therefore, $0\leq E(M_j)\leq 1$.
\end{proof}

\textit{Example.}
Suppose a model receives normalized efficiency scores $\{0.90,0.85,0.80,0.95\}$ for latency, memory footprint, FLOPs, and throughput, respectively. Then $E(M)=\frac{0.90+0.85+0.80+0.95}{4}=0.875$. Therefore, efficiency remains directly comparable with $P(M)$, $CDC(M)$, and $RR(M)$ on the same normalized scale.

\end{document}